\documentstyle[prl,aps,multicol,epsf]{revtex}
\twocolumn
\begin{document}
\draft
\title{ Geometric phase shift in
quantum computation using superconducting nanocircuits: 
nonadiabatic effects}
\author{Shi-Liang Zhu$^{1,2}$ and 
Z. D. Wang$^{1,3}$ \thanks{To whom correspondence should be addressed.
Email address: zwang@hkucc.hku.hk}
}
\address{ 
$^{1}$Department of Physics, University of Hong Kong, Pokfulam Road,
Hong Kong, China\\
$^{2}$Department of Physics, South China Normal University,
Guangzhou, China\\
$^{3}$ Department of Material Science and Engineering, University of 
Science and Technology of China, Hefei, China
}
\address{\mbox{}}
\address{\parbox{14cm}{\rm \mbox{}\mbox{}
The nonadiabatic geometric quantum computation may be achieved
using coupled low-capacitance Josephson juctions.
We show that the nonadiabtic effects
as well as the adiabatic condition
are very important for these systems.
Moreover, we find that 
it may be hard to detect the adiabatic Berry's phase
in this kind of
superconducting nanocircuits; but the nonadiabatic
phase may be measurable with current techniques.
Our results may provide useful information for the implementation of
geometric quantum computation.
}}
\address{\mbox{}}
\address{\parbox{14cm}{\rm PACS numbers:
03.67.Lx, 85.25.Cp, 03.65.Bz,73.23.-b}}
\maketitle

\newpage
\narrowtext

Quantum computation is now attracting increasing interest both 
theoretically and experimentally.
So far, a number of systems have been proposed as potentially
viable quantum computer models,  including
trapped ions, 
cavity quantum electrodynamics,
nuclear magnetic resonce(NMR), {\it etc}~\cite{Cirac}.
In particular, a kind of solid state qubits
using controllable low-capacitance Josephson juctions
has been paid considerable 
attention\cite{Shnirman,Makhlin,Nakamura,Falci}.
A two-qubit gate in many experimental implementations is
the controlled phase shift, which may be achieved using either a
conditional dynamic or geometric phase. A remarkable feature of the 
latter
lies in that it depends only on the geometry of the path 
executed\cite{Berry},
and therefore provides a possibility to perform quantum gate
operations by an intrinsically
fault-tolerant way\cite{Zanardi,Jones}.

Recently, several basic ideas of adiabatic geometric
quantum computation by using NMR\cite{Jones}, 
superconducting nanocircuits\cite{Falci}
or trapped ions\cite{Duan}
were proposed. However, since some of the quantum gates are
quite sensitive to perturbations of the phase factor of the
computational basis states, control 
of the phase factor becomes an important issue
for both hardware and software.
Moreover, the adiabatic evolution appears to be quite special,
and thus the nonadiabatic correction on  the phase shift
may need to be considered in some realistic systems
as it may play a significant role in a whole process~ 
\cite{Aharonov,Zhu,Zhu2}.
In this paper,
we focus on the nonadiabatic 
geometric phase in superconducting nanocircuits.
We indicate that the adiabatic Berry's phase,
as well as the single qubit gate
controlled by this phase,
may  hardly  be implemented in the present experimental
setup.
On the other hand,
since the 2-qubit operations are about $10^2$ times slower
than the 1-bit operations\cite{Makhlin},
the conditional
adiabatic phase 
is extremely difficult to
be achieved.
A serious disadvantange of the adiabatic
conditional phase shift is that the
adiabatic condition
requires that the evolution time must be
much longer than the typical operation time
$\tau_0$ ($=\hbar/E_J$ with $E_J$ as the Josephson energy),
which leads to an intrinsical time limitation
on the operation of quantum gate.
Therefore, a generalization to nonadiabatic cases
is  important in controlling the quantum gates.
We find that the nonadiabatic geometric phase shift
can also be used to achieve the phase shift in quantum gates.

\begin{figure}
\label{fig1}
\epsfxsize=7.5cm
\epsfbox{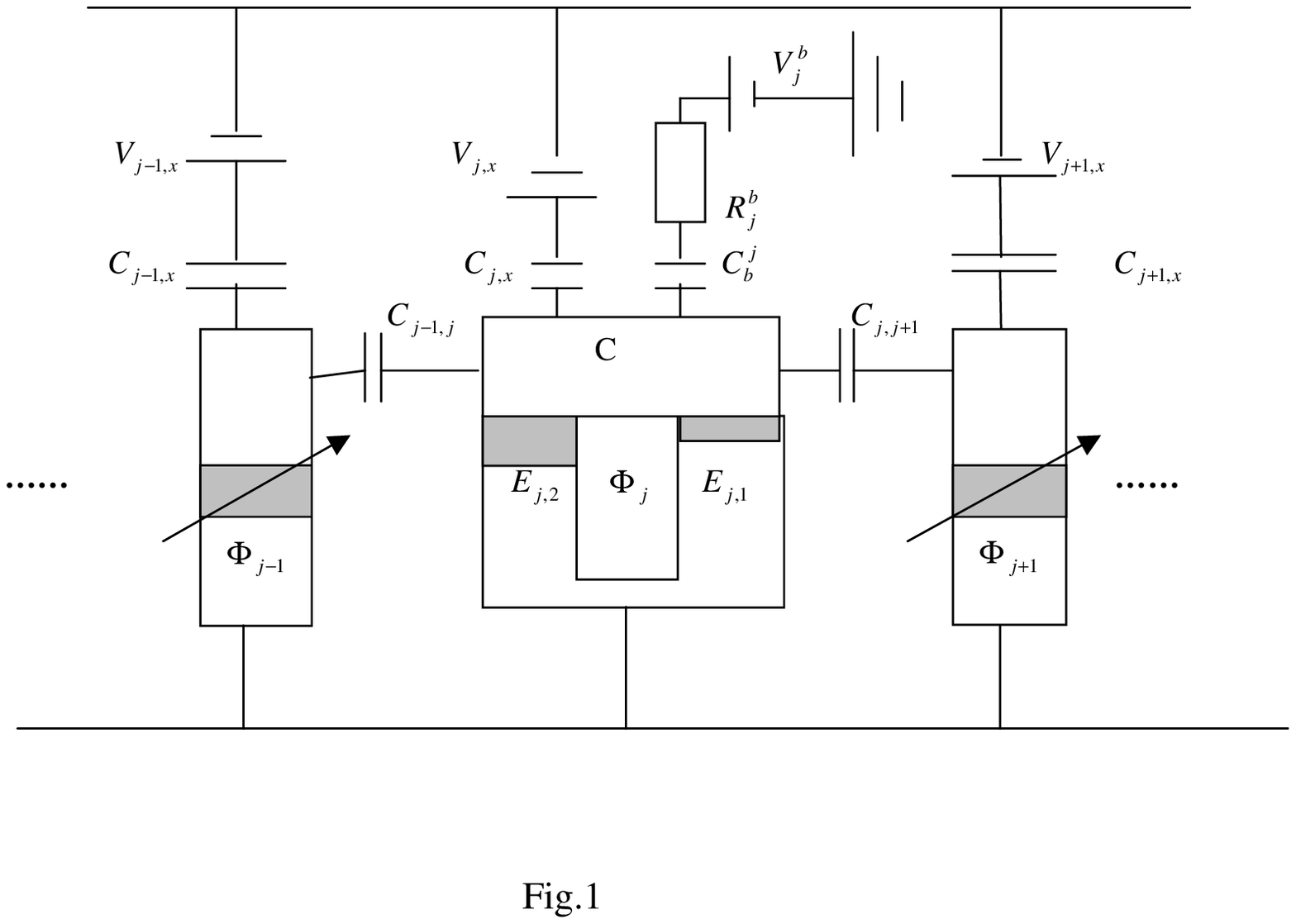}
\vspace{-3.5cm}
\caption{Schematic diagram of a quantum computer. The $j$-th qubit and 
its probe circuit
are displayed in detail.}
\end{figure}

We first consider a single qubit using Josephson juctions
described in Ref.\cite{Falci} ( see the {\sl j}-th qubit in Fig.1).
The qubit  consists of a superconducting
electron box formed by an asymmetric SQUID with
the Josephson coupling $E_{1}$ and $E_{2}$,
pierced by a magnetic flux $\Phi$ and subject to an applied
gate voltage $V_x=2en^e_x/C_x$ (here we omit the subscript $j$,
and $2en^e_x$ is the offset charge).
In the charging regime
(where $E_{1,2}$ are much smaller than the charging energy $E_{ch}$)
and at low temperatures,
the system behaves as an artificial
spin-$1/2$ particle in a magnetic field, and the effective Hamiltonian
reads~\cite{Averin2} 
\begin{equation}
\label{Hamiltonian}
\hat{H}=-\frac{1}{2}{\bf B}\cdot
\stackrel{\rightarrow}{\sigma},
\end{equation}
where $\sigma_{x,y,z}$ are Pauli matrices,
and the fictitious field
\begin{equation}
\label{fictitious}
{\bf B}=\{ E_Jcos\alpha,-E_Jsin\alpha,E_{ch}(1-2n^e_x) \}
\end{equation}
with $E_J=\sqrt{(E_{1}-E_{2})^2+4E_{1}E_{2}cos^2(\pi\Phi/\Phi_0)}$,
$tan\alpha=(E_{1}-E_{2})tan(\pi\Phi/\Phi_0)/(E_{1}+E_{2})$, and
$\Phi_0=h/2e$.
In this qubit Hamiltonian, charging energy is equivalent to
the $B_z$ field whereas the Josephson term determines the fields in
the $xy$ plane. 
By changing $V_x$ and $\Phi$ the qubit Hamiltonian describes a
curve in the parameter space
$\{ {\bf B} \}$.
Therefore by adiabatically changing $\hat{H}$
around a circuit in $\{ {\bf B} \}$,
the eigenstates will accumulate a Berry's phase
$\gamma_B=\mp \Omega/2$, where the
signs $\pm$ depend on whether the system is in
the eigenstate aligned with or against the
field\cite{Berry}. The solid angle $\Omega$, which represents
the magnetic field
trajectory subtends at ${\bf B}=0$,
is derived as
\begin{equation}
\label{solid-angle}
\Omega=\int_0^\tau \frac{B_x\partial_t B_y-B_y\partial_t B_x}
{|{\bf B}|(B_z+|{\bf B}|)}dt,
\end{equation}
under the condition ${\bf B}(\tau)={\bf B}(0)$.

However, the adiabatic evolution is quite special, and thus
the generalization
to nonadiabatic noncyclic cases is of significance. 
We now recall how to
calculate the Pancharatnam phase.
For a spin-$1/2$ particle subject to an arbitrary magnetic field,
each spin state
$|\psi\rangle=
[e^{-i\varphi/2}cos(\theta/2),\ e^{i\varphi/2}sin(\theta/2)]^T$
may be mapped into
a unit vector
${\bf n}=(sin\theta cos\varphi,
sin\theta sin\varphi,cos\theta)$,
with ${\bf n}\in$ a unit sphere $S^2$,
via the relation
${\bf n}=\langle\psi|\stackrel{\rightarrow}{\sigma|}\psi\rangle$,
where $T$ represents the transposition of matrix.
By changing the magnetic field, the evolution of spin state
is a curve on $S^2$
from an initial state $(\theta_i,\varphi_i)$
to a final state $(\theta_f,\varphi_f)$, and
the Pancharatnam phase accumulated in this
evolution was found to be~\cite{Zhu}
\begin{equation}
\label{phase}
\gamma= - \frac{1}{2}\int_{C}
(1-cos\theta)d\varphi
+ arctan\frac {sin(\varphi_f-\varphi_i)}
{cot\frac {\theta_f}{2} cot\frac 
{\theta_i}{2}+cos(\varphi_f-\varphi_i)},
\end{equation}
where $C$ is along the actual evolution curve on
$S^2$, and is determined by
the equation: $\partial_t{\bf n}(t)=
-{\bf B}(t)\times{\bf n}(t)/\hbar$.
This $\gamma$ phase recovers the Aharonov-Anandan (AA)
phase (Berry's phase) in a cyclic
(adiabatic)
evolution\cite{Zhu}.

At this stage, we propose how to
detect the nonadiabatic or adiabatic geometric phase in
the charge qubit system.
The system is prepared in the ground state of the Hamiltonian at
$n^e_x=0$ and $\Phi=0$, and then
changes to the fictitious field ${\bf B}(\Phi(t), n_x^e(t))$,
which is a periodic function of time $t$ with the period
$\tau$. We consider the process where a pair of
orthogonal states
$|\psi_\pm\rangle$ evolve cyclically(but not necessary adiabatically).
This process can be realized in the present system.
Noting that the adiabatic approximation is merely a sufficient but not
 necessary condition for the above cyclic
evolution,
we here focus on a nonadiabatic generalization.
In this evolution, the initial state is given by
$$
|\psi_i\rangle=a_{+}|\psi_{+}(\theta_i,\varphi_i)\rangle
+a_{-}|\psi_{-}(\theta_i,\varphi_i)\rangle,
$$
where
\begin{eqnarray*}
&|&\psi_{+}(\theta,\varphi)\rangle
=[e^{-i\varphi/2}cos(\theta/2),\ e^{i\varphi/2}sin(\theta/2)]^T,\\
&|&\psi_{-}(\theta,\varphi)\rangle
=[-e^{-i\varphi/2}sin(\theta/2),\ e^{i\varphi/2}cos(\theta/2)]^T,\\
&a&_{+}=cos[(\eta-\theta_i)/2] cos\varphi_i/2
-i cos[(\eta+\theta_i)/2] sin\varphi_i/2,\\
&a&_{-}=sin[(\eta-\theta_i)/2] cos\varphi_i/2
+i sin[(\eta+\theta_i)/2] sin\varphi_i/2,
\end{eqnarray*}
with $tan\eta=E_J(\Phi=0)/E_{ch}$,
$tan\theta_i=[E_J(t)/B_z(t)]|_{t=0}$,
and $tan\varphi_i=[B_y(t)/B_x(t)]|_{t=0}$.
A phase difference between $|\psi_{+/-}\rangle$
can be introduced by changing $\hat{H}$. The phases acquired in this 
way will have both  geometrical and  dynamical components. But
the dynamical phase accumulated in the whole procedure
may be removed\cite{Dynamic},
thus only the geometric phase
remains. By taking into account the cyclic condition
${\bf n}(0)={\bf n}(\tau)$ for $|\psi_\pm\rangle$,
the final state in this case
is given by\cite{Note}
\begin{equation}
\label{final}
|\psi_f\rangle=a_{+}e^{i\gamma}
|\psi_{+}(\theta_i,\varphi_i)\rangle
+a_{-}e^{-i\gamma}
|\psi_{-}(\theta_i,\varphi_i)\rangle,
\end{equation}
where $\gamma$ can be calculated from Eq.(\ref{phase}).
The contribution from the second term of
Eq.(\ref{phase}) vanishes simply because ${\bf n}(0)={\bf n}(\tau)$.
Thus the geometric phase considered here is the cyclic
AA phase.
The probability of measuring a charge
$2e$ ($n=1$) in the box at the end of this procedure is
derived as
\begin{equation}
\label{prob}
P_1=|a_{+}sin\frac{\theta_i}{2}+a_{-}cos\frac{\theta_i}{2} 
e^{-2i\gamma}|^2.
\end{equation}
This probability can be simplified to
\begin{equation}
\label{probability}
P_1=[1-cos(\eta-\theta_i) cos\theta_i+sin(\eta-\theta_i)
sin\theta_i cos2\gamma]/2.
\end{equation}
when $\Phi(0)=0$.
Note that Eq.(\ref{probability})
recovers $sin^2(\gamma)$ in Ref.\cite{Falci}
even in a nonadiabatic but cyclic evolution\cite{Cyclic}. 
Thus the nonadiabatic phase
may be determined by
the probability of the charge state
in the box at the end of this process.
It is worth pointing out that the parameters
$\eta$ and $\theta_i$ in Eq.(\ref{prob})(or(\ref{probability}))
are fully determined by
the experimentally controllable parameters
$\Phi$ and $n_x^e$, as in
the adiabatic Berry's phase case~\cite{Falci}.

It is remarkable that the probability
obtained in Eq.(\ref{prob})(or(\ref{probability}))
may be directly
detected by
the dc current through the probe junction $C_{b}^j$  
under a finite bias voltage $V_{b}^j$\cite{Nakamura}.
Assume that we have achieved
one SQUID qubit as well as
the detector circuit, as shown in Fig.1.
By changing $V_{jx}$ and $\Phi_j$
in time $[0,\tau]$,
the system oscillates between $|0\rangle$ and $|1\rangle$, and the 
final
state would be determined by the geometric phase. The measurable dc
current through the probe junction formulates by the processes: 
$|1\rangle$
emits two electrons to the probe,
while $|0\rangle$ does nothing.
Consequently, the probability described
by Eq.(\ref{prob})(or (\ref{probability})) as well as
the geometric phase may be detected by the dc current.

The single qubit gate
may be realized by this geometric phase. For example,
it is straightforward to check that the unitary evolution
operator defined by $|\psi_f\rangle=U^{sq}_1|\psi_i\rangle$,
is given by
\begin{equation}
\label{single-qubit}
U_{1}^{sq}(\gamma)=\left (
\begin{array}{ll}
cos\gamma & isin\gamma  \\
isin\gamma   &  cos\gamma
\end{array}
\right ),
\end{equation}
when $\theta_i=0$ and $\varphi_i=0$.
Clearly, the operation depends on the geometric phase 
$\gamma$; 
$\gamma=\pi/2$ and  $\gamma=\pi/4$
procduce a spin
flip (NOT-operation) and an equal-weight 
superposition of spin states, respectively.
On the other hand, the phase-flip gate
$U_2^{sq}=exp(-2i\gamma|1\rangle\langle 1|)$
(up to an irrelevant over phase ) is derived by
$\theta_i=0$ and $\varphi_i=0$.
The noncommutable $U_1^{sq}$ and $U_2^{sq}$
gates are the two well-known universal gates for
single-qubit operation.
The Berry's phase may be used to achieve
intrinsical fault-tolerant quantum computation
since it depends only on the
evolution path  in the parameter space.
The nonadiabatic cyclic phase is also rather universal 
in a sense that it is the same for a infinite number
of possible ways of motion  along the curves in the
projective Hilbert space\cite{Aharonov}.
Consequently, the nonadiabatic phase may also
be used as a tool for
some fault-tolerant quantum computation.

We now illustrate how to achieve the cyclic
state for quantum gates in two processes.
The parameters $(\Phi(t),n_x^e(t))$ in  process I change as
\begin{equation}
\label{procedure1}
(I) \left \{
\begin{array}{ll}
\{ \frac{4\Phi_m t}{\tau},\frac{1}{2} \}, & t\in [0,\frac{\tau}{4})  \\
\{ 
\Phi_m,\frac{1}{2}+4(n_{xm}^e-\frac{1}{2})(\frac{t}{\tau}-\frac{1}{4}) \},
& t\in [\frac{\tau}{4},\frac{1\tau}{2})  \\
\{ -4\Phi_m t/\tau+3\Phi_m,n_{xm}^e \}, & t\in 
[\frac{1\tau}{2}, \frac{3\tau}{4}) \\
\{ 0, n_{xm}+4(\frac{1}{2}-n_{xm}^e)(\frac{t}{\tau}-\frac{3}{4}) \}.& 
t\in [\frac{3\tau}{4},\tau) 
\end{array}
\right.
\end{equation}
The path in the parameter space $\{ {\bf B} \}$
swept out in this case is exactly the same as that proposed 
in Ref.\cite{Falci}, 
Since the evolution in  this process 
is cyclic only under the adiabatic condition,
we need to answer a key question: whether
the adiabatic approximation is valid for 
the given parameters?
As for process II ,
the parameters $(\Phi(t),n_x^e(t))$  change as
\begin{equation}
\label{procedure2}
(II)
\left\{
\begin{array}{c}
\Phi(t)=\frac{\Phi_0}{\pi}atan[\frac{E1+E2}{E1-E2}tan(\omega t)],\\
n^e_x(t)=\frac{1}{2}(1-\frac{E_Jctg\chi_0+\hbar\omega}{E_{ch}}).
\end{array}
\right.
\end{equation}
The fictitious field described by Eq.(\ref{procedure2})
guarantees that the angle
$\chi_0=arctan[E_J/(B_z(t)-\hbar\omega)]$ (and $n_z$)
is time-independent. It is found that the state described by the
vectors ${\bf n}(\chi_0,-\omega t)$ in this process evolves
cyclically  with period $\tau=2\pi/\omega$ \cite{Wang},
and the AA phase for one cycle
is given by $\gamma=\pi(1-cos\chi_0)$,
which may be used to achieve the mentioned
single-qubit gates
geometrically. For the present system,
the dynamic phase can be removed by 
simply choosing
$\omega=-4(E_1+E_2) E_k[-4E_1 E_2/(E_1-E_2)^2]/\pi sin(2\chi_0)$
with $E_k (x)$ the complete elliptic integral of the first kind.

\begin{figure}
\label{fig2}
\epsfxsize=7.5cm
\epsfbox{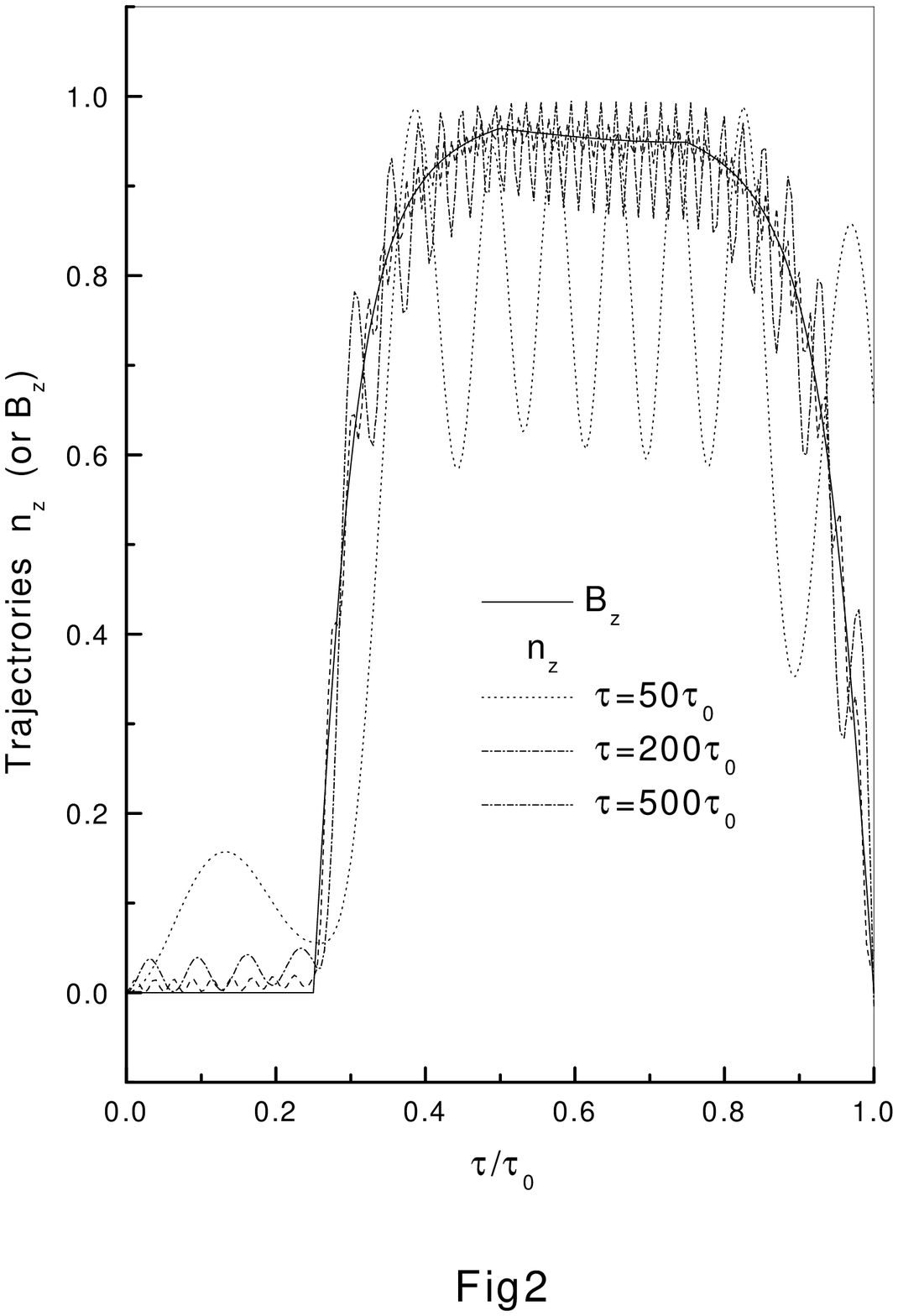}
\vspace{-0.3cm}
\caption{
The trajectories $n_z$ and $\hat{B}_z$ versus time
in  process I for $\Phi_m=0.25$, $n^e_{xm}=0.20$, $E_2=4E_1=6.25\mu ev$,
and $E_{ch}=5.0(E_1+E_2)$.
}
\end{figure}

The nonadiabatic effect should be important
if $\tau$ is not short.
We first consider the evolutions described
by Eq.(\ref{procedure1}).
Figure 2 shows $n_z(t)$ and $\hat{B}_z(t)=B_z(t)/|{\bf B}(t)|$ versus 
time, with the parameters being the same as those in Ref.\cite{Nakamura}.
The deviation of ${\bf n}(t)$ from $\hat{{\bf B}}(t)(={\bf B}(t)/|{\bf
B}(t)|)$ indicates clearly 
whether or not the adiabatic approximation
is valid because   ${\bf n}(t)$ almost
follows the trajectory of
the magnetic field $\hat{{\bf B}}(t)$ under this approximation.
It is seen from Fig.2 
that the adiabatic approximation  is satisfied in the first case
when $\tau>500\tau_0$,
where $\tau_0=\hbar/(E_1+E_2)\sim 84ps$.
The adiabatic condition for  process II
is in the same order of magnitude
(see Fig.3).
It is worth pointing out that the coherence time
achieved in a single SQUID is
merely about $30\sim40 \tau_0$\cite{Nakamura},
which is not long enough for the 
adiabatic evolution, implying that the adiabatic condition is
not satisfied in the above two  processes for realistic systems.
But, fortunately,  the nonadiabatic phase
can be measured and used in achieving geometric quantum gates.

Conditional geometric phase
accumulated in one sub-system evolution depends
on the quantum state of another sub-system,
which may be realized by coupling capacitively two
asymmetric SQUIDS (see any neighboring pair of 
qubits in Fig.1.).
If the coupling capacitance $C_{ij}$ is smaller than the others,
the Hamiltonian reads
\begin{equation}
\label{coupling}
\hat{H}=\sum\limits_{i=1}^{N}\hat{H}_i
+\sum\limits_{i=1}^{N-1}(\hat{H}_{i,i+1}+H.C.),
\end{equation}
where $H_i$ refer to the uncoupled qubits defined in
Eq.(\ref{Hamiltonian}) and $\hat{H}_{i,i+1}
=E_{i,i+1}(n^e_i-n^e_{x,i})(n^e_{i+1}-n^e_{x,i+1})$
with $E_{i,i+1}=E_{ch}C_{i,i+1}/C$\cite{Falci}.
The gate voltage and magnetic flux
can be independently fixed for all qubits.
We address firstly a two-qubit operation,
e.g., $i$ and $j$ qubits are two
neighbour qubits with the {\sl i}-th as the control qubit and
the {\sl j}-th as the target qubit.
The fictitious field on the target qubit
is $[E_J(\Phi_j)cos\alpha_j,-E_J(\Phi_j)sin\alpha_j,
B_z^l]$ with $B_z^l=
E_{ch}(1-2n^e_{x,j})+E_{i,j}(n^e_{x,i}-l)$,
where $l$ represent the control qubit state
$0$ or $1$.
Obviously, the geometric phase
$\gamma_j$ for $j$-th qubit in decoupled case
is different from $\gamma_j^l$
even the changeings of $(\Phi_j,n^e_{x,j})$
are the same,
where $\gamma_j^l$ is the geometric phase of the target qubit
when the charge state of the control qubit is
$l$. $\gamma_j^l$ may be directly derived from
Eq.(\ref{phase}).
It is worth to pointing out that
the state described by the
vector ${\bf n}(\chi^l,-\omega t)$
with $\chi_0^l=arctan[E_J/(B_z^l-\hbar \omega)]$
is still a cyclic evolution, and
may be used to achieve the two-qubit operation.
In terms of the basis $\{|00\rangle,|01\rangle,|10\rangle,
|11\rangle \}$,  the unitary operator to describe
the two-qubit gate is given by~\cite{Falci}
\begin{equation}
\label{unitary}
U_{(\gamma^0_j,\gamma^1_j)}=
diag(e^{-i\gamma^0_j},e^{i\gamma^0_j},e^{-i\gamma^1_j},
e^{i\gamma_j^1}).
\end{equation}
The combination with single-bit operations allow
us to perform the XOR gate. The unitary operation
for XOR gate can be obtained by
$U_{XOR}=\left [
I\otimes U_1^{sq}(\pi/4) \right ]
U_{(0,\frac{3\pi}{2})}
\left [ I\otimes U_1^{sq}(\pi/4) \right ]^\dagger$
with $I$ as a $2\times 2$ unit matrix.
This XOR gate together with single qubit gates constitutes a 
universality:
they are sufficient for all
manipulations required for quantum computation~\cite{Lloyd}.
Therefore, all the elements of quantum computation may be
achievable by (nonadiabatic) geometric phase.

We now compute the geometric phases
required by the spin flip operation
($\gamma=3\pi/2$) and NOT operation
($\gamma=3\pi/4$)
accumulated in the second process.
The comparison of the nonadiabatic geometric phase
$\gamma$ with $\gamma_a$ is shown in Fig.3,
where the $\gamma_a$ is the phase
calculated under the adiabatic approximation.
It is seen that the $\gamma_a$ deviates evidently from
the  $\gamma$ for $\tau<150\tau_0$.
Thus the operation time required by the
adiabatic condition in
both processes I and II is in the same order of magnitude.
The dynamic phases can be removed
when $\tau \sim 3.57\tau_0$
for $\gamma=3\pi/2$ and
$\tau\sim 4.12\tau_0$ for
$\gamma=3\pi/4$, respecitively.
Therefore, by accurately controlling
the parameters $\Phi$ and $n_x^e$,
we may control the state in the projective Hilbert space.
It is striking that the present operation time for
nonadiabatic geometric gates is much shorter than that 
 with the adiabatic scheme.
Note that
the coherence time
achieved in a single SQUID by current
technology is
about $30\sim40 \tau_0$\cite{Nakamura},
implying that tens of geometric NOT
operation may be achieved experimentally.
Therefore, the generalization of the adiabatic phase to the
nonadiabatic case is of significance
since the coherence time achieved
in charge qubit in Josephson Junctions
is short.
Moreover, the large number qubits
required for useful computation may be devised
by a network similar to Fig.1.

\begin{figure}
\label{fig3}
\epsfxsize=7.5cm
\epsfbox{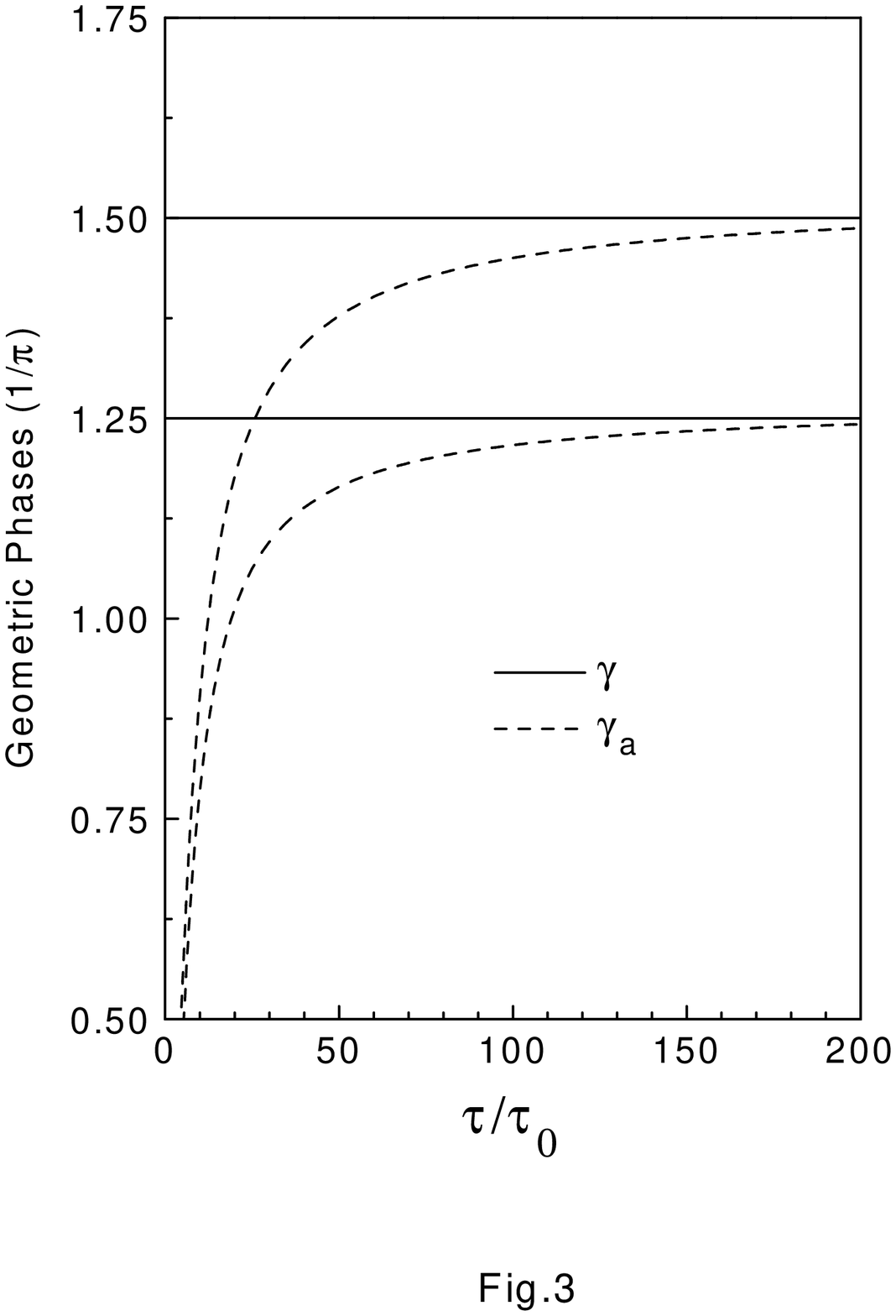}
\vspace{-0.3cm}
\caption{The comparison of adiabatic phases
with nonadiabatic ones.
The parameters are the same as those in Fig.2.}
\end{figure}

In conclusion, we study how to detect the nonadiabatic phase
in superconducting nanocircuits, and the possibility to use
the nonadiabatic phase as a tool to achieve the
quantum computation.

We wish to acknowledge valuable discussions with
Dr. L. M. Duan and Dr. X. B. Wang. This work was supported  by
the RGC grant of Hong Kong
under Grant Nos. HKU7118/00P and HKU7114/02P,
the Ministry of Science and Technology of
China under Grant No. G1999064602, and the URC fund of HKU.
S. L. Z. was supported in part by the SRF for ROCS, SEM,
the NSF of Guangdong under Grant No. 021088, and the NNSF
of China under Grant No. 10204008.


\begin{references}
\bibitem{Dynamic} 
A useful method is to choose some specific external parameters
such that the dynamical phase accumulated in the whole process
is zero(also see Ref.[17b]).
\bibitem{Cyclic} 
The initial state for the procedure described
in Ref.\cite{Falci}
is specially chosen as 
$\theta_i=\pi/2$.  $(E_J(\Phi=0)/E_{ch})^2$ is also neglected.
\end{references}
\end{document}